# A THz Video SAR Imaging Algorithm Based on Chirp Scaling

Jiawei Jiang
School of Optoelectronic Information and Computer Engineering
University of Shanghai for Science and Technology
Shanghai, China
202310315@st.usst.edu.cn

Yinwei Li *
Terahertz Technology Innovation Research Institute
University of Shanghai for Science and Technology
Shanghai, China
liyw@usst.edu.cn

Qibin Zheng
Terahertz Technology Innovation Research Institute
University of Shanghai for Science and Technology
Shanghai, China
qbzheng@usst.edu.cn

*Abstract*—In video synthetic aperture radar (SAR) imaging mode, the polar format algorithm (PFA) is more computational effective than the backprojection algorithm (BPA). However, the two-dimensional (2-D) interpolation in PFA greatly affects its computational speed, which is detrimental to the real-time imaging of video SAR. In this paper, a terahertz (THz) video SAR imaging algorithm based on chirp scaling is proposed, which utilizes the small synthetic angular feature of THz SAR and the inherent property of linear frequency modulation. Then, two-step chirp scaling is used to replace the 2-D interpolation in the PFA to obtain a similar focusing effect, but with a faster operation. Point target simulation is used to verify the effectiveness of the proposed method.

## I. INTRODUCTION

The concept of video SAR is first proposed by Sandia National Laboratories (SNL) in 2003. In video SAR imaging mode, the radar system moves around the target scene and provides continuous surveillance then displays the received information in a movie-like format [1]. Compared with traditional SAR, video SAR can observe the target in all time and all weather, and its ability to generate continuous frame images can detect changes in the target scene more intuitively. To achieve this video-like effect, the frame rate of video SAR needs to exceed 5 Hz. According to the principle of spotlight SAR imaging, the frame rate of video SAR is proportional to the carrier frequency at a certain azimuth resolution [2]. Therefore, terahertz band (0.1-10 THz) is an excellent choice for video SAR. In addition, high frame rates have demanding requirements on the efficiency of video SAR imaging algorithms, so it is critical to select imaging algorithms with high computational efficiency.

The difference in imaging geometry makes some of the imaging algorithms applicable to stripmap SAR no longer applicable to circular SAR (CSAR). The two most commonly used imaging algorithms in CSAR are backprojection algorithm (BPA) and polar format algorithm (PFA). BPA do not use any approximation and can achieve high resolution imaging under arbitrary flight trajectory, but its huge number of operations leads to long imaging time [3]. PFA simplifies the imaging process by using the planar wavefront assumption and has a higher operation speed than BPA, which makes it more suitable for real-time imaging of video SAR. However, the 2-D interpolation from polar coordinates to rectangular coordinates in PFA greatly affects its computational speed [4], which is unfavorable for high frame rate imaging of video SAR. The alternative methods to interpolation include chirp-Z transform (CZT) [5] and chirp scaling [6]. However, chirp scaling can achieve a higher operation rate than CZT. In this paper, a THz video SAR imaging algorithm based on chirp scaling is proposed. The method decomposes the 2-D interpolation into range and azimuth interpolation by using the small synthetic angle property of THz SAR. Next, two-step chirp scaling is used to replace two-step interpolation using the inherent property of linear frequency modulation (LFM).This method involves only the fast Fourier transform (FFT) and complex multiplication, which improves the computational speed of algorithm and is more conducive to processing on hardware such as FPGA.

## II. VIDEO SAR GEOMETRIC MODEL

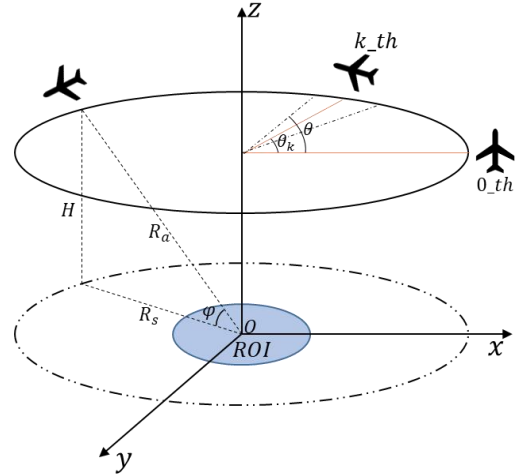

Fig. 1. Video SAR imaging mode

The imaging geometry model of video SAR is shown in Fig. 1. The radar flies along a circular path of radius $R_s$ in a horizontal plane at a fixed height H with a constant velocity v. The beam illuminates the region of interest (ROI) during the radar flight. Where θ and φ denote the azimuth angle and elevation angle of the radar platform, respectively, $θ_k$ represents the central azimuth of the kth frame sub-aperture, and the variable $R_a = R_s/\cosφ$ denotes the slant range between the antenna phase center (APC) and the scene center O. The position of the APC can be expressed as:

$$\begin{aligned}x_a(t) &= R_a \cosφ \cosθ \\ y_a(t) &= R_a \cosφ \sinθ \\ z_a(t) &= R_a \sinφ\end{aligned} \quad (1)$$

The transmission pulse should be LFM for an effective utilization of chirp scaling. The dechirped signal of an arbitrary ground target after removing the residual video phase and envelope skew terms takes the form:

$$s(\tau, t) = \sigma \cdot \text{rect}\left[\frac{\tau - \Delta t}{T_r}\right] \exp\left[-j\frac{4\pi}{c}(f_c + K\tau)\Delta R\right] \quad (2)$$

where $\sigma$ denotes the scattering intensity of the target, $\tau$ is the fast time, $\Delta t = 2R_t/c$ denotes the time delay for a point target with distance to the radar of $R_t$. $T_r$ is the pulse width, $f_c$ is the center frequency, and K denotes the chirp rate of the LFM. $\Delta R$ is the difference range of the point target with respect to the reference point, which can be expressed as:

$$\Delta R = \sqrt{[x_a(t) - x]^2 + [y_a(t) - y]^2 + [z_a(t)]^2} - R_a \quad (3)$$

Using the planar wavefront assumption, the difference range $\Delta R$ can be expanded to:

$$\Delta R \approx -(x \cos\varphi \cos\theta + y \cos\varphi \sin\theta) \quad (4)$$

Next, let $K_X$ and $K_Y$ denote the range and azimuth wavenumber components, respectively:

$$\begin{cases} K_X = \dfrac{4\pi}{c}(f_c + K\tau)\cos\varphi\cos\theta \\ K_Y = \dfrac{4\pi}{c}(f_c + K\tau)\cos\varphi\sin\theta \end{cases} \quad (5)$$

Then (2) can be expressed in the spatial wavenumber domain as follows:

$$S(K_X, K_Y) = \sigma \cdot \text{rect}\left[\frac{\tau - \Delta t}{T_r}\right] \exp[j(xK_X + yK_Y)] \quad (6)$$

In the conventional PFA, the conversion from wavenumber domain signals to focused SAR images can be achieved by a 2-D Fourier transform, which can be realized by an efficient 2D-FFT. However, the FFT requires uniformly spaced samples on a rectangular grid. As shown in (5), the samples after mapping to the spatial wavenumber domain $(K_X, K_Y)$ are distributed on a polar grid, so the 2D resampling in the spatial frequency domain is required to make them uniformly distributed on a rectangular grid before applying the 2D-FFT. The process is shown in Fig. 2.

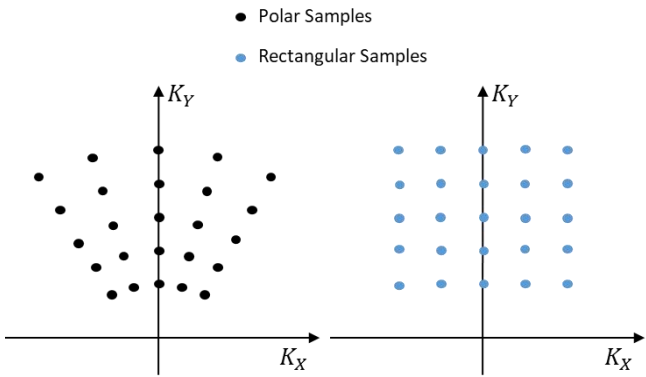

Fig. 2. 2-D resampling from polar coordinates to rectangular coordinates.

### III. VIDEO SAR IMAGE ALGORITHM BASED ON CHIRP SCALING

The process of 2-D resampling in the traditional PFA is achieved by 2-D interpolation, which is usually considered as an inefficient method in the field of SAR signal processing. However, the LFM has a special property that enables a non-interpolation conversion from polar coordinates to rectangular coordinates. Multiplying LFM with another LFM with the same chirp rate, the result is still LFM, but with a small shift in phase center and chirp rate, this property of LFM can be called chirp scaling [7]. Due to the small synthesis angle characteristic of video SAR in terahertz band [8], 2-D resampling can be separated into two consecutive 1-D resampling, which are realized by range and azimuth chirp scaling, respectively. An interpolation-free video SAR imaging algorithm is achieved by this method.

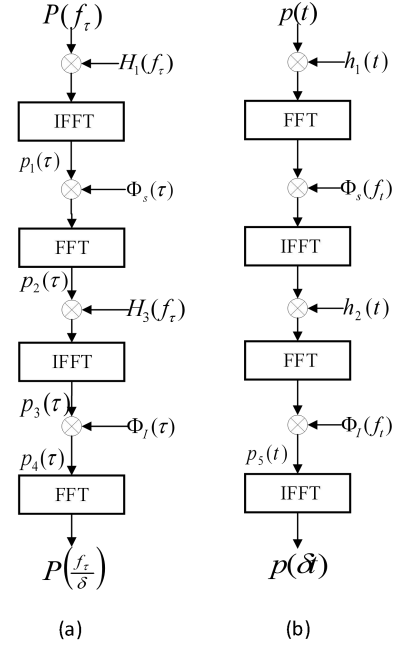

Fig. 3. (a) Frequency scaling for range resampling (b) Time scaling for azimuth resampling

Fig. 3 demonstrates the flowchart of time scaling and frequency scaling for PFA respectively, where the frequency scaling is derived on the basis of time scaling using the convolution property of the Fourier transform. Fig. 3(a) represents the process of range resampling, the input signal $P(f_\tau) = \mathbf{F}_r[s_R(\tau, t)]$, where $\mathbf{F}_r[\cdot]$ represents the range FFT, and $s_R(\tau, t)$ represents the received dechirped signal.

According to the property of Fourier transform, the signal $p_1(\tau)$ can be expressed as a convolution of $p(\tau)$ with $h_1(\tau) = \exp(j\alpha\tau^2)$:

$$p_1(\tau) = \exp(j\alpha\tau^2)\int_{-\infty}^{\infty} p(u)\exp[j\alpha(u^2 - 2\tau u)]\,du \quad (7)$$

Define the system function of range scaling as:

$$H_1(f_\tau) = \exp\left(-j\frac{\pi^2 f_\tau^2}{\alpha}\right) = \exp\left(j\frac{\pi f_\tau^2}{K}\right) \quad (8)$$

It is obvious that $\alpha = -\pi K$. Then the received dechirped signal can be obtained at $p_1(\tau)$:

$$p_1(\tau) = \mathbf{F}_r^{-1}\{\mathbf{F}_r[s(\tau, t)] \cdot H_1(f_\tau)\} = s_R(\tau, t) \quad (9)$$

where $\mathbf{F}_r^{-1}$ represents the inverse range FFT, the range scaling function is defined as:

$$\Phi_S(\tau) = \exp[j\pi K(1 - \delta_r)(\tau - t_0)^2] \quad (10)$$

where $t_0 = 2R_a/c$ and $\delta_r$ is the range scaling factor:

$$\delta_r = \frac{1}{\cos(\theta - \theta_k)} \quad (11)$$

The product of $p_1(\tau)$ and the range scaling function $\Phi_S(\tau)$ is:

$$p_2(\tau) = \exp\left\{j\alpha\left[\delta_r\tau^2 - (1-\delta_r)[t_0^2 - 2t_0\tau]\right]\right\} \quad (12)$$

The system function for range scaling is defined as:

$$H_2(f_\tau) = \exp\left(-j\frac{\pi f_\tau^2}{\delta_r K}\right)\exp\left\{j2\pi f_\tau\left[\frac{(\delta_r-1)f_c}{\delta_r K}\right]\right\} \quad (13)$$

where $f_\tau$ is the range spectrum.

Then the signal at $p_3(\tau)$ is:

$$p_3(\tau) = \exp\left\{j\pi K\delta_r(1-\delta_r)\left[\tau + \frac{(\delta_r-1)f_c}{\delta_r K} - t_0\right]\right\} \cdot s_R\left[\delta_r(\tau-t_0) + \frac{(\delta_r-1)f_c}{K} + t_0, t\right] \quad (14)$$

Define the inverse range scaling function $\Phi_I(\tau)$ as:

$$\Phi_I(\tau) = \exp\left\{j\pi K\delta_r(\delta_r-1)\left[\tau + \frac{(\delta_r-1)f_c}{\delta_r K} - t_0\right]\right\} \quad (15)$$

Then the range scaled signal can be obtained at $p_4(\tau)$:

$$p_4(\tau) = s_R\left[\delta_r(\tau-t_0) + \frac{(\delta_r-1)f_c}{K} + t_0, t\right] \quad (16)$$

Therefore, the process of range scaling actually starts from $p_1(\tau)$ and ends at $p_4(\tau)$. Moreover, the elimination of RVP and range resampling are completed during the process of range scaling. Fig. 3(b) represents the process of azimuth resampling, where the input signal of azimuth scaling is the output signal $p_4(\tau)$ of range scaling. Correspondingly, the system parameters of azimuth scaling are:

$$h_1(t) = \exp[j\pi K_a t^2] \quad (17)$$

$$\Phi_S(f_t) = \exp\left[j\pi\frac{(\delta_a-1)f_t^2}{\delta_a K_a}\right] \quad (18)$$

$$h_2(t) = \exp[j\pi\delta_a K_a t^2] \quad (19)$$

$$\Phi_I(f_t) = \exp\left[-j\pi\frac{(\delta_a-1)f_t^2}{\delta_a^2 K_a}\right] \quad (20)$$

Where $f_t$ is the frequency variable corresponding to azimuth time t, $\delta_a$ and $K_a$ denote the azimuth scaling factor and the Doppler rate at the aperture center, respectively:

$$\delta_a = \frac{f_c}{f_c + K(\tau-t_0)} \quad (21)$$

$$K_a = \frac{-2v^2}{\lambda R_a} \quad (22)$$

The 2-D focused image can be obtained by doing a 2-D FFT on the azimuth scaled signal. But it is helpful to find that the signal at $p_5(t)$ is azimuth resampled and focused:

$$p_5(t) = \mathbf{F}_a[s_R(\tau, \delta_a t)] \quad (23)$$

where $\mathbf{F}_a[\cdot]$ denotes the azimuth FFT. It is obvious that the azimuth FFT is canceled by the subsequent inverse FFT(IFFT) at $p_5(t)$. Therefore, a range FFT will convert the signal $p_5(t)$ into the 2-D focused video SAR image.

Finally, the flowchart of modified PFA applicable to THz video SAR is shown in Fig. 4.

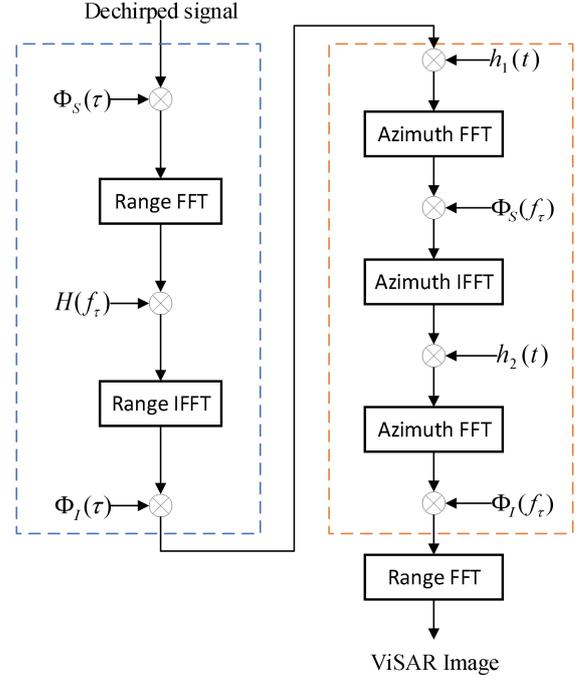

Fig. 4. Flowchart of THz video SAR imaging algorithm using chirp scaling.

IV. POINT TARGET SIMULATION

In this section, point target simulation is employed to verify the effectiveness of the proposed algorithm. The simulation parameters are listed in Table I. The location of the point target is shown in Fig. 5.

TABLE I. SIMULATION PARAMETERS

| Parameters | Values |
|---|---|
| Carrier frequency | 220GHz |
| Bandwidth | 1.2GHz |
| Sampling frequency | 13MHz |
| Pulse width | 80us |
| Pulse repetition frequency | 6kHz |
| Slant Range | 2500m |
| Grazing angle | 45° |

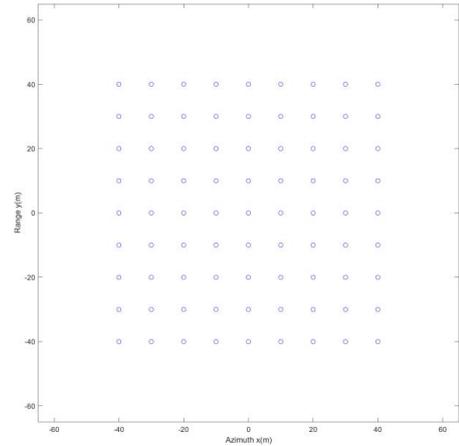

Fig. 5. Distribution of point targets.

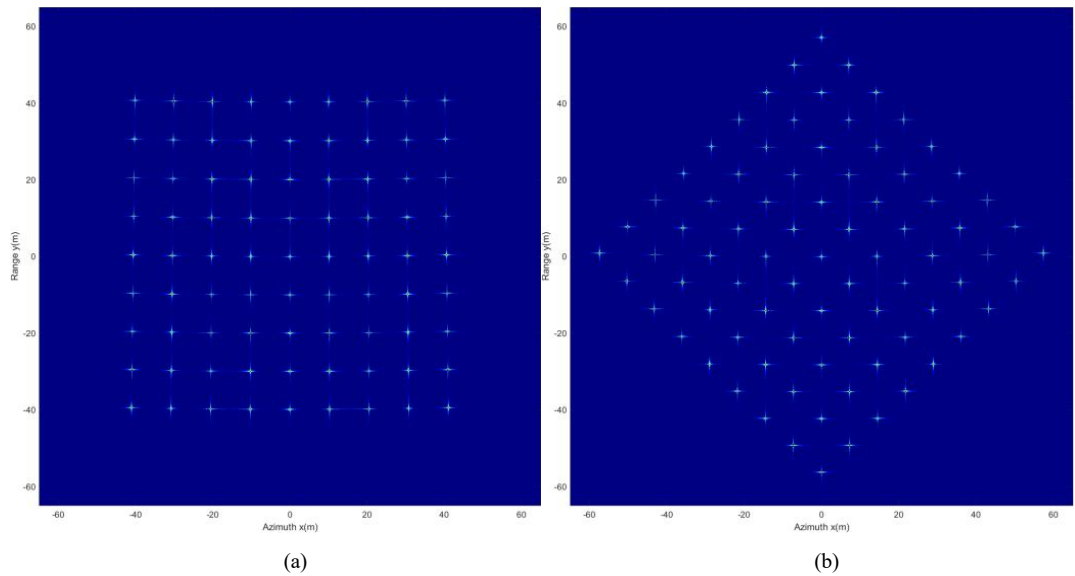

Fig. 6. Imaging results of traditional PFA for different azimuth angles. (a) $\theta_k = 0$ (b) $\theta_k = \pi/4$.

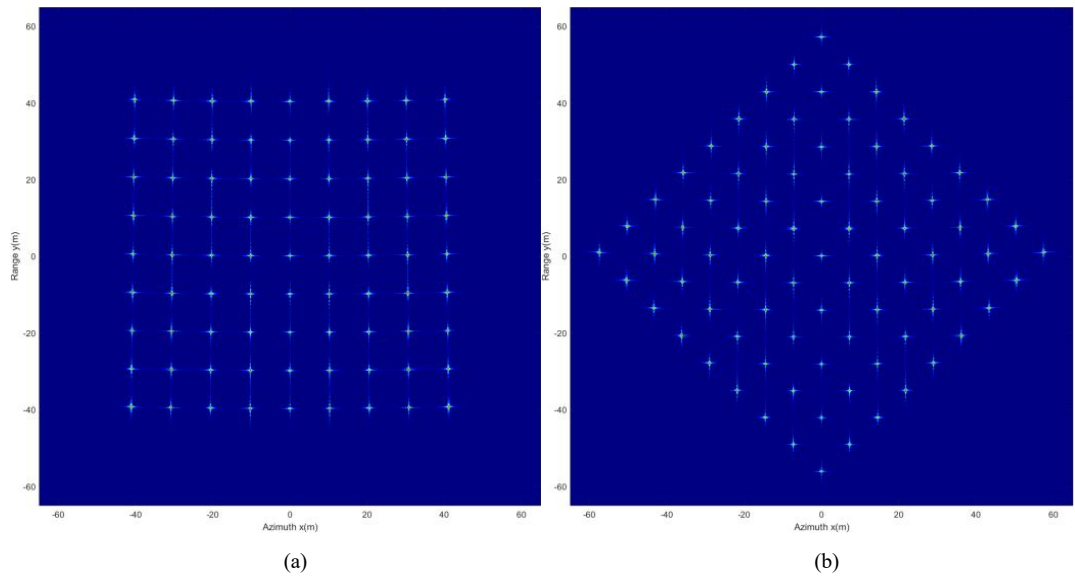

Fig. 7. Imaging results of our method for different azimuth angles. (a) $\theta_k = 0$ (b) $\theta_k = \pi/4$

Fig. 6 and Fig. 7 demonstrate the imaging results of traditional PFA and the proposed algorithm when the azimuth angle is $\theta_k = 0$ and $\theta_k = \pi/4$, respectively. In order to observe the shape of the point spread function (PSF) more intuitively, no windows are performed during the imaging process. It can be seen that the main flap of PSF is parallel to the image coordinate axis and present in a "cross" shape. The point target is well-focused and located in true position, the rotation angle of image is equal to the radar azimuth angle $\theta_k$.

To analyze the imaging quality of traditional PFA and the proposed algorithm, several main parameters of the point target located at (0,0) are listed in Table II. The impulse response width (IRW) specifies the resolution of SAR images, and it can be seen that both methods achieve the theoretical resolution of 0.17 m. In addition, the peak side lobe ratio (PSLR) and the integrated side lobe ratio (ISLR) indicate that the imaging quality of both methods is similar.

TABLE II. IMAGE QUALITIES OF DIFFERENT ALGORITHM

|  | Traditional PFA | | Our method | |
| --- | --- | --- | --- | --- |
|  | Azimuth | Range | Azimuth | Range |
| IRW(m) | 0.1584 | 0.1575 | 0.1599 | 0.1610 |
| PSLR(dB) | -13.2808 | -13.2085 | -13.1705 | -13.1206 |
| ISLR(dB) | -24.9043 | -24.4433 | -24.9997 | -25.1716 |